\documentclass[10pt]{article}
\usepackage{amsmath,amssymb, bm}
\usepackage{graphicx}
\usepackage{hyperref}
\DeclareGraphicsExtensions{.eps,.bmp,.wmf,.jpg,.pdf}
\numberwithin{equation}{section}
\def\be{\begin{equation}}
\def\ee{\end{equation}}

\def\bea{\begin{eqnarray}}
\def\eea{\end{eqnarray}}
\title{\textbf{Cosmological dynamics and observational constraints on a viable $f(Q)$ non-metric gravity model}}
\author{A. Oliveros\thanks{alexanderoliveros@mail.uniatlantico.edu.co}\,\, and
Mario A. Acero\thanks{marioacero@mail.uniatlantico.edu.co}\\
Programa de F\'isica, Universidad del Atl\'antico, Carrera 30 N\'umero 8-49\\
Puerto Colombia-Atl\'antico, Colombia} 

\date{}
\begin{document}
\maketitle

\begin{abstract}
\noindent Inspired by an exponential $f(R)$ gravity model studied in the literature, in this work we introduce a new and viable $f(Q)$ gravity model, which can be represented as a perturbation of $\Lambda$CDM. Typically, within the realm of $f(Q)$ gravity, the customary approach to investigate cosmological evolution involves employing a parametrization of the Hubble expansion rate in terms of the redshift, $H(z)$, among other strategies. In this work  we have implemented a different strategy, deriving an analytical approximation for $H(z)$, from which we deduce approximated analytical expressions for the parameters $w_{\rm{DE}}$, $w_{\rm{eff}}$, and $\Omega_{\rm{DE}}$, as well as the deceleration parameter $q$. In order to verify the viability of this approximate analytical solution, we examined the behavior of the these parameters in the late-time regime\textbf, in terms of the free parameter of the model, $b$. We find that for $b>0$, $w_{\rm{DE}}$ shows a quintessence-like behavior, while for $b<0$, it shows a phantom-like behavior. However, regardless of the sign of $b$, $w_{\rm{eff}}$ exhibits a quintessence-like behavior. Furthermore, it has been deduced that as the magnitude of the parameter $b$ increases, the present model deviates progressively from $\Lambda$CDM. We have also performed a Markov Chain Monte Carlo statistical analysis to test the model predictions with the Hubble parameter, the Pantheon supernova (SN) observational data, and the combination of those samples, obtaining constraints on the parameters of the model and the current values of the Hubble parameter and the matter density. Our findings indicate that this $f(Q)$ gravity model is indeed a viable candidate for describing the late-time evolution of the Universe at the background level.
\end{abstract}


\section{Introduction}\label{sec_intro}
\noindent For over two decades, significant efforts have been dedicated to cosmological research in pursuit of an explanation of the observed late-time cosmic acceleration. The primary avenue of investigation involves the introduction of a novel energy component within the Universe, termed dark energy (DE), which is distinguished by its negative pressure. Nevertheless, as of the present time, a definitive and satisfactory resolution to the enigma of DE remains elusive; its incorporation into the framework of fundamental physics theories continues to challenge researchers (for a comprehensive review about this topic see Refs.~\cite{peebles,copeland,bamba1}). 

One of the intriguing approaches for elucidating the late-time cosmic acceleration, 
beyond the inclusion of either DE or novel forms of matter to account for this phenomenon, lies within the field of modified gravity theories (see e.g.~Refs.~\cite{Sotiriou:2008rp,odintsov1,clifton,Capozziello:2011et,odintsov2,Odintsov:2023weg} for a review). Usually, in this framework, the fundamental action is built assuming generalized functions of the scalar curvature (the so-called $f(R)$ theories), general higher-order theories, scalar–tensor theories of gravitation, etc. Recently, a new proposal has emerged within the realm of modified theories of gravitation. 
These particular theories, in which gravitational interactions are governed by non-metricity, with curvature and torsion being rendered negligible, 
are known as $f(Q)$ theories or $f(Q)$ symmetric teleparallel gravity, where $Q$ is the non-metricity scalar \cite{nester, BeltranJimenez, lavinia1, lavinia2, runkla, Bahamonde:2021gfp}. These theoretical frameworks hold the potential to provide fresh perspectives on the phenomenon of cosmic acceleration, stemming from the inherent consequences of an alternative geometry as opposed to the conventional Riemannian framework (see Ref.~\cite{lavinia} for a recent and extensive review on this topic). 

Although this proposal is very recent, there are numerous works in the literature that have been carried out using it \cite{lazkoz1, barros, mandal1, mandal2, khyllep1, saridakis, atayde, dambrosio, solanki, narawade1, capozziello, arora, khyllep2, lymperis, koussour1, koussour2, gadbail, DAgostino:2022tdk, de, boehmer, narawade2, sokoliuk, andronikos1, myrzakulov1, ferreira, mussatayeva, myrzakulov2, pooja, aguiar, mandal, koussour3, andronikos2, aoki, iosifidis, nojiri}. Usually, owing to the presence of nonlinear elements within the field equations, one of the principal challenges inherent in these scenarios pertains to the task of deriving solutions, whether through analytical or numerical means. Although, commonly the field equations are solved numerically, it is also an usual 
try to propose a parametrization of either the Hubble parameter, the equation of state parameter or $f(Q)$ in terms of the redshift, among other strategies. 

For instance, in Ref.~\cite{lazkoz1} the authors performed an observational analysis of several modified $f(Q)$ models using the redshift approach, where the $f(Q)$ Lagrangian is reformulated as an explicit function of the redshift, $f(z)$. Various different polynomial parameterizations of $f(z)$ are proposed, including new terms which would allow for deviations from the $\Lambda$CDM model. In Ref.~\cite{koussour1} a new parametrization of the Hubble parameter is proposed in a model-independent way and apply it to the Friedmann equations in the FLRW Universe. Also, the authors of Ref.~\cite{myrzakulov2} implemented a parametrization scheme for the Hubble parameter, obtaining an exact solution for the field equations in $f(Q)$ cosmology.

The proposal presented here is a new viable $f(Q)$ gravity model which can be represented as a perturbation of the $\Lambda$CDM model. This model is inspired by an exponential $f(R)$ gravity model studied recently in the literature \cite{granda, oliveros1, oliveros}, which depends upon two parameters, $b$ and $n$. Furthermore, instead of using the aforementioned strategies, we have chosen to implement the formalism developed in Ref.~\cite{basilakos} within the context of $f(R)$ gravity. This formalism involves expressing the solution to the field equations, i.e., $H^2(N)$, as a perturbative series expansion in terms of the deviation parameter $b$ (making $n=1$), with 
the $\Lambda$CDM model solution as a reference. Additionally, we deduce approximated analytical expressions for the parameters $w_{\rm{DE}}$, $w_{\rm{eff}}$, and $\Omega_{\rm{DE}}$, as well as the deceleration parameter $q$ in terms of the redshift $z$. Finally, in order to verify the viability of this approximate analytical solution for $H(z)$ in late-times, we examine the behavior of these cosmological parameters, and perform a fit to observational data of the Hubble parameter ($H(z)_{{\rm obs}}$) and the apparent magnitude ($m_b(z)_{{\rm obs}}$, Pantheon SN sample) in terms of the redshift, to obtain constraints on the model parameter $b$, together with the present values of the expansion rate, $H_0$, and the matter density parameter, $\Omega_{m0}$.

This paper is organized as follows: In Section \ref{model}, we provide a concise overview of $f(Q)$ gravity. In Section \ref{cosmo-analysis}, we introduce the $f(Q)$ gravity model, which plays the central role in this work, 
and carry out 
a perturbative expansion in the parameter $b$. This expansion leads to an analytic expression for $H(\Omega_{m0},b;z)$ at all orders in $b$. Additionally, we deduce approximated analytical expressions for the parameters $w_{\rm{DE}}$, $w_{\rm{eff}}$, and $\Omega_{\rm{DE}}$, as well as the deceleration parameter $q$ in terms of the redshift $z$, and we also plot them to investigate their evolution. Section \ref{params} is devoted to establishing constraints to the parameters involved in the model, considering two sets of observational data, and our conclusions are presented in Section \ref{conclus}.
\section{The $f(Q)$ gravity: a brief review}\label{model} 
\noindent In general, the action for an $f(Q)$ gravity model in the presence of matter components is given by
\begin{equation}\label{eq_action}
S = \int{d^4x \sqrt{-g} \left(\frac{f(Q)}{2\kappa^2} + \mathcal{L}_{\rm{M}}\right)},
\end{equation}
where $g$ denotes the determinant of the metric tensor, $g^{\mu\nu}$, $\kappa^2=8\pi G=1/M_{\rm{p}}^2$, with $G$ being the Newton's constant, and $M_{\rm{p}}$ the reduced Planck mass. $\mathcal{L}_{\rm{M}}$ represents the Lagrangian density for the matter components (relativistic and non-relativistic perfect matter fluids). The term $f(Q)$ is for now an arbitrary function of the non-metricity scalar $Q$. This scalar it is obtained by the contraction of the non-metricity tensor, which is defined as \cite{lavinia, lavinia1, lavinia2}
\begin{equation}\label{eq_Qdef}
Q_{\gamma\mu\nu} \equiv \nabla_\gamma g_{\mu\nu} = \partial_\gamma g_{\mu\nu} - \Gamma^{\beta}_{\mkern10mu\gamma\mu}g_{\beta\nu} - \Gamma^{\beta}_{\mkern10mu\gamma\nu}g_{\beta\mu},
\end{equation}
where $\nabla_\gamma$ represents the covariant derivative with respect to a general affine connection $\Gamma^{\alpha}_{\mkern10mu\gamma\beta}$, which can be written as \cite{hehl, ortin}
\begin{equation}\label{eq_affine}
\Gamma^\lambda\\
_{\mu\nu}=\left\{^\lambda\\
_{\mu\nu}\right\}+K^\lambda\\
_{\mu\nu}+L^\lambda\\
_{\mu\nu},
\end{equation}
with the Levi-Civita connection of the metric given by
\begin{equation}\label{eq_LeviCivita}
\left\{^\lambda\\
_{\mu\nu}\right\}\equiv\frac{1}{2}g^{\lambda\beta}\left(\partial_\mu g_{\beta\nu}+\partial_\nu g_{\beta\mu}-\partial_\beta g_{\mu\nu}\right);
\end{equation}
the contortion tensor is
\begin{equation}\label{eq_contortion}
K^\lambda\\
_{\mu\nu}\equiv\frac{1}{2}T^\lambda\\
_{\mu\nu}+T^{\,\,\,\,\,\lambda}_{(\mu\,\,\,\,\nu)},
\end{equation}
with the torsion tensor $T^\lambda_{\,\,\,\,\mu\nu} \equiv 2\Gamma^\lambda_{\,\,\,\,[\mu\nu]}$, and the disformation tensor $L^\lambda_{\,\,\,\,\mu\nu}$, which can be written in terms of the non-metricity tensor as
\begin{equation}\label{eq4}
L^{\beta}_{\mkern10mu\mu\nu}=\frac{1}{2}Q^{\beta}_{\mkern10mu\mu\nu}-Q_{(\mu\nu)}^{\mkern35mu\beta}.
\end{equation}
The contraction of the tensor $Q_{\gamma\mu\nu}$ is given by
\begin{equation}\label{eq2}
Q = -Q_{\gamma\mu\nu} P^{\gamma\mu\nu},
\end{equation}
where the tensor $P^{\gamma\mu\nu}$ is the non-metricity conjugate, defined as
\begin{equation}\label{eq_metricityConj}
P^{\beta}_{\mkern10mu\mu\nu}=-\frac{1}{2}L^{\beta}_{\mkern10mu\mu\nu}+\frac{1}{4}(Q^\beta-\tilde{Q}^\beta)g_{\mu\nu}
-\frac{1}{4}\delta^{\beta}_{\,\,\,(\mu}Q_{\nu)},
\end{equation}
the other quantities present in Eq.~(\ref{eq_metricityConj}) are given by
\begin{equation}\label{eq5}
Q_{\beta}=g^{\mu\nu}Q_{\beta\mu\nu},\quad \tilde{Q}_{\beta}=g^{\mu\nu}Q_{\mu\beta\nu}.
\end{equation}
Variation of the action (\ref{eq_action}) with respect to the metric gives the equation of motion \cite{lavinia1, lavinia2},
\begin{equation}\label{eq_MotionEq}
\frac{2}{\sqrt{-g}}\nabla_\beta(f_Q\sqrt{-g}P^\beta_{\mkern10mu\mu\nu}) + \frac{1}{2}fg_{\mu\nu} + f_Q(P_{\mu\beta\lambda}Q_\nu^{\mkern10mu\beta\lambda}-2Q_{\beta\lambda\mu}P^{\beta\lambda}\mkern1mu\nu) = -T_{\mu\nu},
\end{equation}
where $f_Q \equiv \frac{df}{dQ}$, and as usual, the energy momentum tensor $T_{\mu\nu}$ is given by
\begin{equation}\label{eq_Tmunu}
T_{\mu\nu}=-\frac{2}{\sqrt{-g}}\frac{\delta\sqrt{-g}\mathcal{L}_{\rm{M}}}{\delta g^{\mu\nu}},
\end{equation}
which in this case is considered to be a perfect fluid, i.e.,~$T_{\mu\nu}=(\rho+p)u_\mu u_\nu+pg_{\mu\nu}$, where $p$ and $\rho$ are the total pressure and total energy density of any perfect fluid of matter, radiation and DE, respectively. The quantity $u^\mu$ represents the 4-velocity of the fluid.

Considering the flat Friedman-Robertson-Walker (FRW) metric,
\begin{equation}\label{eq_FRWmetric}
ds^2 = -dt^2 + a(t)^2\delta_{ij}dx^idx^j,
\end{equation}
with $a(t)$ representing the scale factor, the time and spatial components of Eq. (\ref{eq_MotionEq}) are given, respectively, by \cite{lavinia1, lavinia2}
\begin{equation}\label{eq_timeEq}
6f_QH^2 - \frac{1}{2}f = \rho_{\rm{m}} + \rho_{\rm{r}},
\end{equation}
and 
\begin{equation}\label{eq_spaceEq}
(12H^2f_{QQ} + f_{Q})\dot{H} = -\frac{1}{2}\left(\rho_{\rm{m}} + \frac{4}{3}\rho_{\rm{r}}\right),
\end{equation}
where $\rho_{\rm{m}}$ is the matter density and $\rho_{\rm{r}}$ denotes the density of radiation. $f_{QQ} = \frac{d^2f}{dQ^2}$, the over-dot denotes a derivative with respect to the cosmic time $t$ and $H\equiv \dot{a}/a$ is the Hubble parameter. Additionally, we have considered $\kappa^2 = 1$. Further, the non-metricity scalar associated with the metric (\ref{eq_FRWmetric}) is given by
\begin{equation}\label{eq_Qscalar}
Q=6H^2.
\end{equation}
It should be noted that in this scenario, the standard Friedmann equations of General Relativity (GR) plus the cosmological constant are recovered when assuming that $f(Q)=Q+2\Lambda$.

If there is no interaction between non-relativistic matter and radiation, then these components obey separately the conservation laws
\begin{equation}\label{eq11}
\dot{\rho}_{\rm{m}} + 3H\rho_{\rm{m}} = 0,\quad \dot{\rho}_{\rm{r}} + 4H\rho_{\rm{r}} = 0.
\end{equation}
As usual in the literature, it is possible to rewrite the field equations (\ref{eq_timeEq}) and (\ref{eq_spaceEq}) in the Einstein-Hilbert form:
\begin{equation}\label{eq12}
3H^2 = \rho,
\end{equation}
\begin{equation}\label{eq13}
-2\dot{H}^2 = \rho + p,
\end{equation}
where $\rho = \rho_{\rm{m}}+ \rho_{\rm{r}} + \rho_{\rm{DE}}$ and $p = p_{\rm{m}} + p_{\rm{r}} + p_{\rm{DE}}$ correspond to the total effective energy pressure densities of the cosmological fluid, respectively.  In this case, the dark energy component has a geometric origin, and after some manipulation in Eqs.~(\ref{eq_timeEq})  and (\ref{eq_spaceEq}), we obtain the effective dark energy and pressure corresponding to the $f(Q)$-theory given by
\begin{equation}\label{eq_effRhoDE}
\rho_{\rm{DE}} = \frac{1}{2}f + 3H^2(1-2f_Q),
\end{equation}
and
\begin{equation}\label{eq_effPDE}
p_{\rm{DE}} = 2\dot{H}(12H^2f_{QQ} + f_Q - 1) - \rho_{\rm{DE}}.
\end{equation}
It is easy to show that $\rho_{\rm{DE}}$ and $p_{\rm{DE}}$ defined in this way satisfy the usual energy conservation equation
\begin{equation}\label{eq16}
\dot{\rho}_{\rm{DE}}+3H(\rho_{\rm{DE}}+p_{\rm{DE}})=0.
\end{equation}
In this case, we assume that the equation of state parameter for this effective dark energy satisfies the following relation:
\begin{equation}\label{eq17}
w_{\rm{DE}} = \frac{p_{\rm{DE}}}{\rho_{\rm{DE}}} = -1 + \frac{2\dot{H}(12H^2f_{QQ} + f_Q - 1)}{\frac{1}{2}f + 3H^2(1 - 2f_Q)}.
\end{equation}
\section{Cosmological dynamics in late-time}\label{cosmo-analysis}
\noindent In this section, we implement the above results taking into account a particular choice for $f(Q)$, and study the resulting late-time cosmological evolution at the background level. To begin with, we introduce the $f(Q)$ gravity model, which plays a central role in this work:
\begin{equation}\label{eq_fQmodel}
f(Q) = Q + 2\Lambda e^{-\left(\frac{b\Lambda}{Q}\right)^n},
\end{equation}
where $\Lambda$ is the cosmological constant, and $b$ and $n$ are real dimensionless parameters. This model is inspired in that studied in Refs.~\cite{granda, oliveros1, oliveros}, but in the context of $f(R)$ gravity. It is evident that for $b=0$ the model given by Eq.~(\ref{eq_fQmodel}) 
is equivalent to GR plus the cosmological constant. In particular, from the structure of this model, it can be seen as a  perturbative deviation from the $\Lambda$CDM Lagrangian. In this sense, this model can be arbitrarily close to $\Lambda$CDM, depending on the parameter $b$, i.e., the exponential form presented in the Eq.~(\ref{eq_fQmodel}) elegantly and succinctly encapsulates the perturbative nature of the model across all orders in the parameter $b$, which is evident from a Taylor series expansion about this parameter,
\begin{equation}\label{eq_fQmodel2}
f(Q) = (Q + 2\Lambda)-\frac{2\Lambda^2 b}{Q}+\frac{\Lambda^3 b^2}{Q^2}-\frac{\Lambda^4 b^3}{3Q^3}+\frac{\Lambda^5 b^4}{12Q^4}-\cdots,
\end{equation}
where for simplicity we have considered $n=1$. It should be highlighted that in the literature also other exponential $f(Q)$ gravity models have been intensively studied (see e.g., Refs.~\cite{saridakis, arora, khyllep2, lymperis, boehmer, narawade2, sokoliuk, ferreira}).\\

Following the procedure carried out in Ref.~\cite{basilakos}, we rewrite Eq.~(\ref{eq_timeEq}) in terms of $N = \ln{a}$
\begin{equation}\label{eq19}
2f_QH^2(N)-\frac{1}{6}f-(\Omega_{m0}e^{-3N}+\Omega_{r0}e^{-4N})H_0^2=0,                    
\end{equation}
where we have introduced the current energy density parameters, $\Omega_{m0} = \frac{\rho_{m0}}{3M_{\rm{p}}^2H_0^2}$ and $\Omega_{r0} = \frac{\rho_{r0}}{3M_{\rm{p}}^2H_0^2}$; here $M_{\rm{p}}=1$, and $H_0$ corresponds to the current Hubble parameter. Implicitly, we consider by simplicity $n=1$ and $\Lambda = 3(1-\Omega_{m0} - \Omega_{r0})H_0^2$. Replacing Eq.~(\ref{eq_fQmodel}) in Eq.~(\ref{eq19}) and taking into account Eq.~(\ref{eq_Qscalar}), we obtain an algebraic equation for $H^2(N)$, given by
\begin{equation}\label{eq19'}
\begin{aligned}
&H^2(N)-(\Omega_{m0}e^{-3N}+\Omega_{r0}e^{-4N})H_0^2\\
&+\frac{e^{\frac{bH_0^2(\Omega_{m0}+\Omega_{r0}-1)}{2H^2(N)}}H_0^2(\Omega_{m0}+\Omega_{r0}-1)[bH_0^2(\Omega_{m0}+\Omega_{r0}-1)+H^2(N)]}{H^2(N)}=0.
\end{aligned}
\end{equation}
Since the above is a transcendental equation, it cannot be solved for $H^2(N)$ in explicit form. Another possibility involves constructing a first-order differential equation for $H^2(N)$ using Eqs.~(\ref{eq_timeEq}) and (\ref{eq_spaceEq}). However, in this scenario the equation is nonlinear and lacks an analytical solution for $H^2(N)$.

Usually, in the literature for circumventing this issue, a common practice is to propose a parametrization of the Hubble parameter in terms of the redshift \cite{lazkoz1, koussour1, myrzakulov2}, among other strategies. In this work instead of use that strategy, we use the formalism developed in Ref.~\cite{basilakos} in the context of $f(R)$ gravity, which consists in expressing the solution to Eq.~(\ref{eq19}), i.e., $H^2(N)$, as a perturbative series expansion in terms of the deviation parameter $b$, with respect to the $\Lambda$CDM model solution as a reference, as follows
\begin{equation}\label{eq20}
H^2(N) = H^2_{\Lambda}(N) + \sum_{i=1}^{M}b^i\delta H_i^2(N),
\end{equation}
where
\begin{equation}\label{eq21}
\frac{H^2_{\Lambda}(N)}{H_0^2} = \Omega_{m0}\,e^{-3N} + \Omega_{r0}\,e^{-4N} + (1 - \Omega_{m0} - \Omega_{r0}) = E_{\Lambda}^2(N),
\end{equation}
further, as has been demonstrate in Ref.~\cite{basilakos}, we can consider only two terms in the above series expansion; in doing so, Eq.~(\ref{eq20}) reduces to
\begin{equation}\label{eq22}
H^2(N) \approx H^2_{\Lambda}(N) + b\,\delta H_1^2(N) + b^2\delta H_2^2(N).
\end{equation}
Replacing Eq.~(\ref{eq22}) in Eq.~(\ref{eq19'}) and considering up to second order in $b$, it is straightforward to calculate $\delta H_1^2(N)$ and $\delta H_2^2(N)$, which are given by
\begin{equation}\label{eq23}
\frac{\delta H_1^2(N)}{H_0^2} = -\frac{3H_0^2(\Omega_{m0} + \Omega_{r0} - 1)^2}{2H_{\Lambda}^{2}(N)},
\end{equation}
\begin{equation}\label{eq24}
\frac{\delta H_2^2(N)}{H_0^2} = -\frac{H_0^4(\Omega_{m0}+\Omega_{r0}-1)^3[18H_0^2(\Omega_{m0}+\Omega_{r0}-1)+5H_{\Lambda}^{2}(N)]}{8(H_{\Lambda}^{2}(N))^3}.
\end{equation}
Now, replacing Eqs.~(\ref{eq23})  and (\ref{eq24}) in Eq.~(\ref{eq22}), and using Eq.~(\ref{eq21}), we obtain an approximate solution for the Hubble parameter $H(z)$:
\begin{equation}\label{eq25}
\begin{aligned}
E^2(z)=&\frac{H^2(z)}{H_0^2}\approx 1-\Omega_{m0}+(1+z)^3\Omega_{m0}-\frac{3(\Omega_{m0}-1)^2}{2[1-\Omega_{m0}+(1+z)^3\Omega_{m0}]}b\\
&+\frac{(\Omega_{m0}-1)^3[(18+5z(3+z(3+z)))\Omega_{m0}-13]}{8[1-\Omega_{m0}+(1+z)^3\Omega_{m0}]^3}b^2,
\end{aligned}
\end{equation}
where for simplicity, we have assumed $\Omega_{r0}=0$ and made the substitution $N=-\ln{(1+z)}$. 

In order to verify the viability of this approximate analytical solution for $H(z)$,  we examine the behaviour of the cosmological parameters $w_{\rm{DE}}$, $w_{\rm{eff}}$, and $\Omega_{\rm{DE}}$, as well as the deceleration parameter $q$ in the late-time regime.  Let us recall that, in terms of the redshift $z$, the cosmological parameters 
are given by
\begin{equation}\label{eq26}
w_{\rm{DE}}=-1+\frac{1}{3}(1+z)\frac{(\rho_{\rm{DE}}(z))'}{\rho_{\rm{DE}}(z)},
\end{equation}
\begin{equation}\label{eq27}
w_{\rm{eff}}=-1+\frac{1}{3}(1+z)\frac{(E^2(z))'}{E^2(z)},
\end{equation}
\begin{equation}\label{eq28}
\Omega_{\rm{DE}}=1-\frac{\Omega_{m0}(1+z)^3}{E^2(z)};
\end{equation}
similarly, the deceleration parameter $q$ 
is given by
\begin{equation}\label{eq29}
q=-1+\frac{1}{2}(1+z)\frac{(E^2(z))'}{E^2(z)},
\end{equation}								
where the prime denotes differentiation with respect to $z$. Using Eqs.~(\ref{eq_effRhoDE}) and (\ref{eq25}) and considering up to second order expansion in $b$, we obtain approximated analytical expressions for the above parameters in terms of the redshift $z$, as follows:
\begin{equation}\label{eq30}
\begin{aligned}
w_{\rm{DE}}\approx &-1-\frac{3[(1+z)^3(\Omega_{m0}-1)\Omega_{m0}]}{2[1-\Omega_{m0}+(1+z)^3\Omega_{m0}]^2}b\\
&+\frac{(1+z)^3(\Omega_{m0}-1)^2\Omega_{m0}[31+(4z(3+z(3+z))-27)\Omega_{m0}]}{4[1-\Omega_{m0}+(1+z)^3\Omega_{m0}]^4}b^2,
\end{aligned}
\end{equation}
\begin{equation}\label{eq31}
\begin{aligned}
w_{\rm{eff}}\approx &-1+\frac{(1+z)^3\Omega_{m0}}{1-\Omega_{m0}+(1+z)^3\Omega_{m0}}+\frac{3(1+z)^3(\Omega_{m0}-1)^2\Omega_{m0}}{[1-\Omega_{m0}+(1+z)^3\Omega_{m0}]^3}b\\
&+\frac{3(1+z)^3(\Omega_{m0}-1)^3\Omega_{m0}[(36+5z(3+z(3+z)))\Omega_{m0}-31]}{8[1-\Omega_{m0}+(1+z)^3\Omega_{m0}]^5}b^2,
\end{aligned}
\end{equation}
\begin{equation}\label{eq32}
\begin{aligned}
\Omega_{\rm{DE}}\approx & 1-\frac{(1+z)^3\Omega_{m0}}{1-\Omega_{m0}+(1+z)^3\Omega_{m0}}-\frac{3(1+z)^3(1-\Omega_{m0})^2\Omega_{m0}}{2[1-\Omega_{m0}+(1+z)^3\Omega_{m0}]^3}b\\
&-\frac{(1+z)^3(\Omega_{m0}-1)^3\Omega_{m0}[(36+5z(3+z(3+z)))\Omega_{m0}-31]}{8[1-\Omega_{m0}+(1+z)^3\Omega_{m0}]^5}b^2,
\end{aligned}
\end{equation}
and
\begin{equation}\label{eq33}
\begin{aligned}
q\approx &-1+\frac{3(1+z)^3\Omega_{m0}}{2[1-\Omega_{m0}+(1+z)^3\Omega_{m0}]}+\frac{9(1+z)^3(\Omega_{m0}-1)^2\Omega_{m0}}{2[1-\Omega_{m0}+(1+z)^3\Omega_{m0}]^3}b\\
&+\frac{9(1+z)^3(\Omega_{m0}-1)^3\Omega_{m0}[(36+5z(3+z(3+z)))\Omega_{m0}-31]}{16[1-\Omega_{m0}+(1+z)^3\Omega_{m0}]^5}b^2.
\end{aligned}
\end{equation}	
Notice that, as it would have been expected, the terms independent of $b$ in each one of the last expressions correspond to those associate to $\Lambda$CDM model.

With Eqs.~(\ref{eq30})-(\ref{eq33}), we can plot the 
evolution of each parameter in terms of the redshift $z$. Additionally, in order to compare the results with the $\Lambda$CDM model, we have also incorporated in the corresponding plots the behavior associated with each quantity defined by Eqs.~(\ref{eq26})-(\ref{eq29}), but using  Eq.~(\ref{eq21}) instead of (\ref{eq25}). 

\begin{figure*}
\centering
    \includegraphics[width=1.0\textwidth]{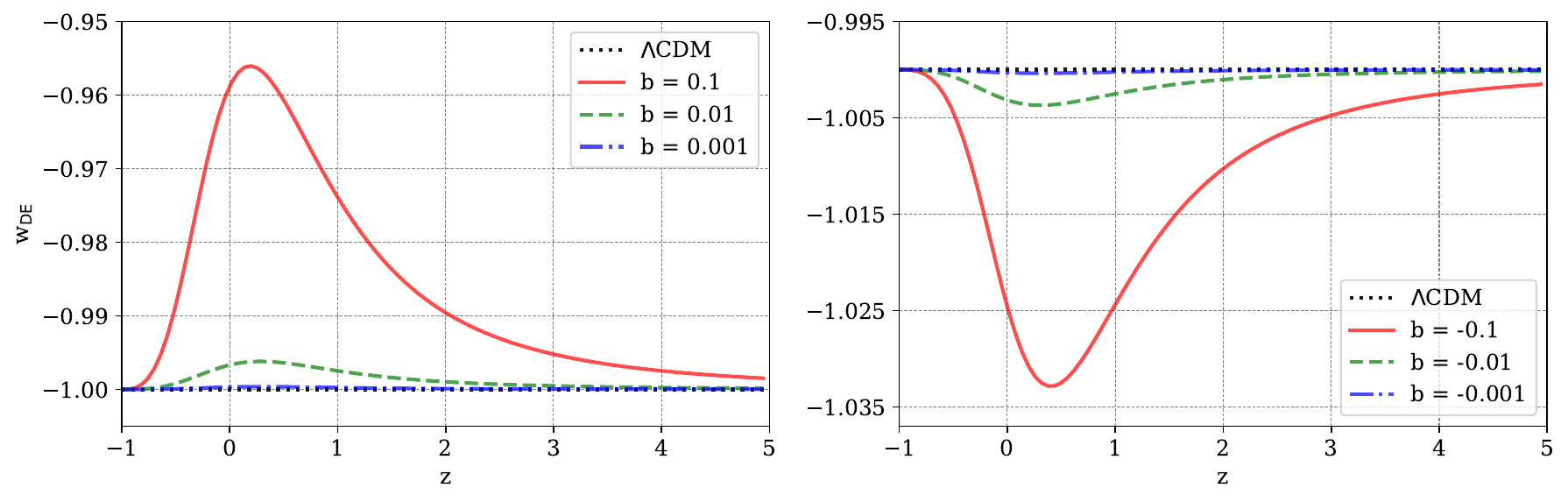}
    \caption{Plot for $w_{\rm{DE}}$ vs. $z$ using positive (\emph{left}) and negative (\emph{right}) values for the parameter $b$. In these plots we have used $\Omega_{m0}=0.315$ }
    \label{fig1}
\end{figure*}

In the left panel of Fig.~\ref{fig1} we show the cosmological evolution of $w_{\rm{DE}}$ as a function of $z$ using some fixed positive values for $b$ and we can see that for large values of $z$ and for the distant future ($z\approx -1$), $w_{\rm{DE}}$ is very close to the $\Lambda$CDM model prediction (i.e. $w_{\Lambda}=-1$). For small values of $z$ (the DE era), $w_{\rm{DE}}$ present a variation about -1, and 
as $b$ decreases, $w_{\rm{DE}}$ progressively converges towards the $\Lambda$CDM model, implying that this variation represents the deviation from $\Lambda$CDM model and it is directly related to the parameter $b$. A similar behavior was found for the $f(R)$ model studied in Ref.~\cite{oliveros}, but in our case the equation of state does not cross the barrier of -1, i.e.~for $b>0$, $w_{\rm{DE}}$ shows a quintessence like behavior. 

In the right panel of Fig.~\ref{fig1}, 
the evolution of the same parameter is shown 
for $b < 0$. 
The same general behavior as for positive $b$ is observed, however $w_{\rm{DE}}$ exhibits now a phantom like behavior, i.e.~$w_{\rm{DE}} < -1$ for all values of $z$. A similar behavior for $w_{\rm{DE}}$ was found in Refs.~\cite{arora} and \cite{lymperis}, but in the context of a different exponential $f(Q)$ theories and a numerical solution of the field equations. 

\begin{figure*}
\centering
    \includegraphics[width=1.0\textwidth]{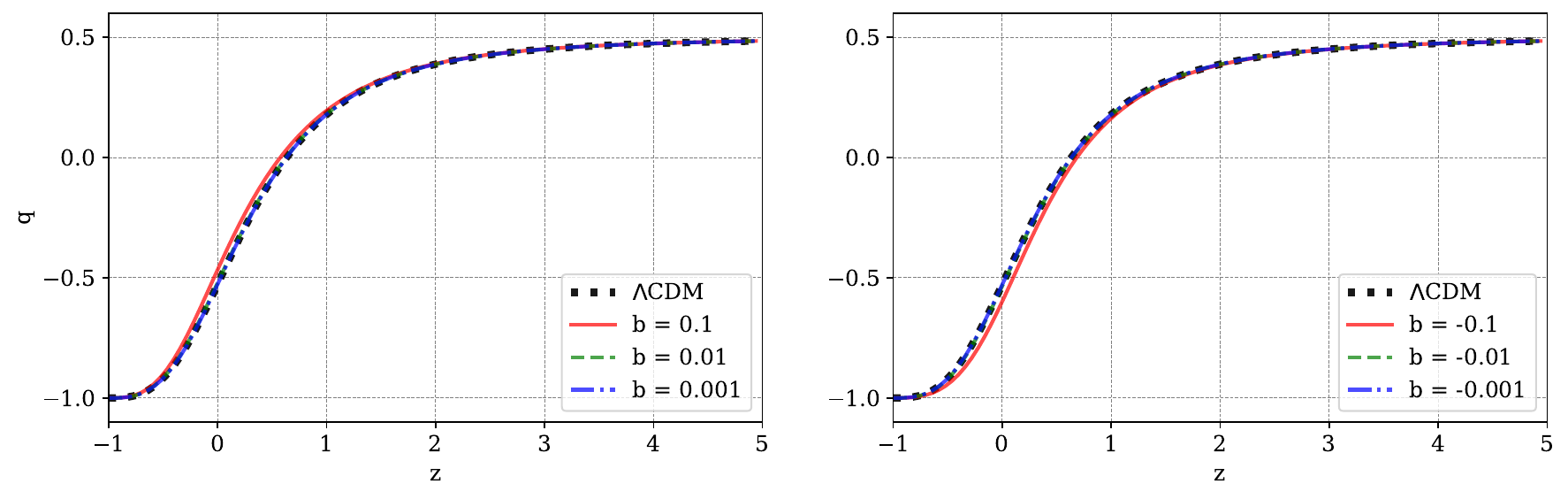}
    \caption{Plot for $q$ vs. $z$ using positive (\emph{left}) and negative (\emph{right}) values for the parameter $b$}
    \label{fig2}
\end{figure*}

\begin{figure*}
\centering
    \includegraphics[width=0.5\textwidth]{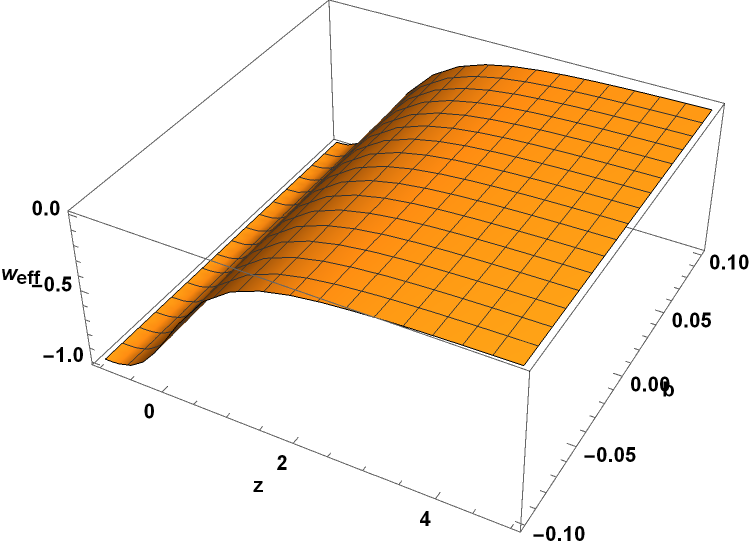}
    \includegraphics[width=0.47\textwidth]{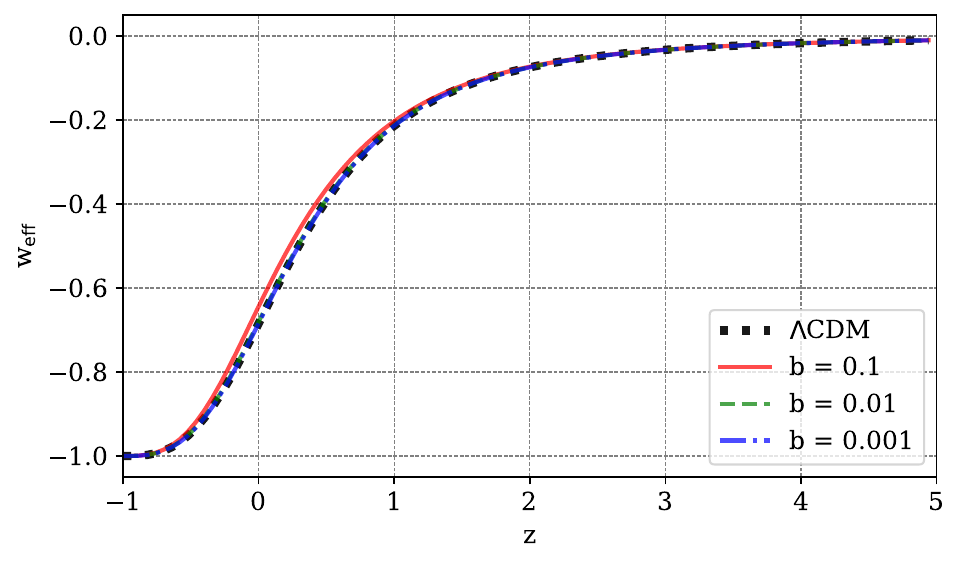}
    \caption{(\emph{left}) Evolution of $w_{\rm{eff}}$ as a function of $z$ and $b$ (positive and negative values) . 
		(\emph{right}) $w_{\rm{eff}}$ vs. $z$.}
    \label{fig3}
\end{figure*}

\begin{figure*}
\centering
    \includegraphics[width=1.0\textwidth]{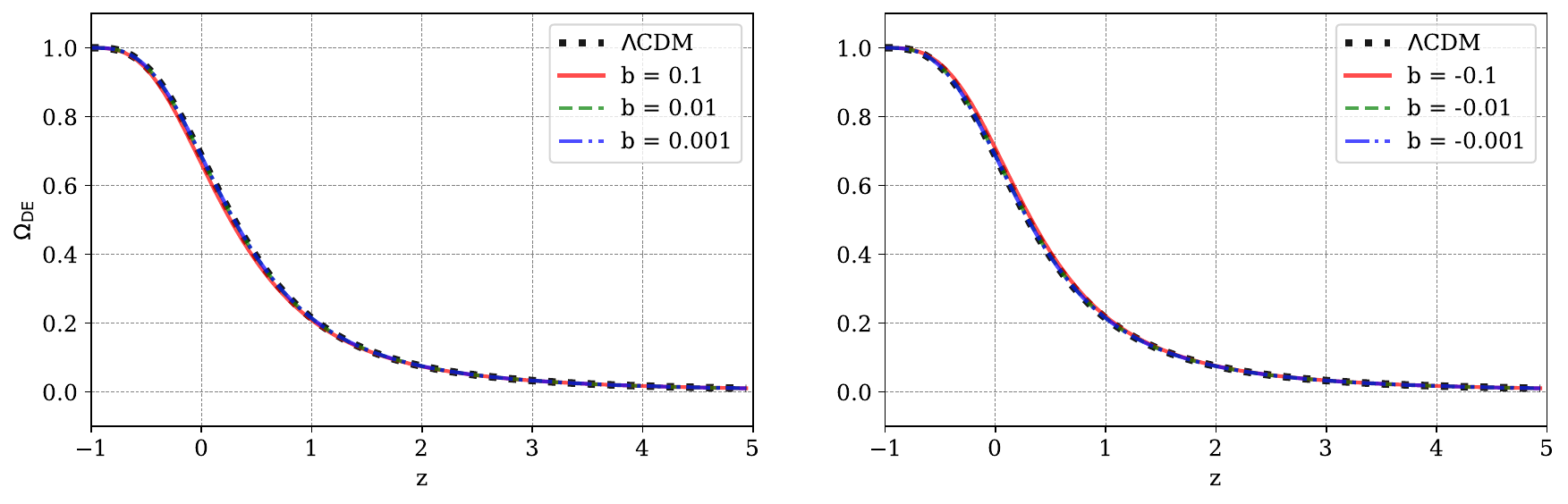}
    \caption{Plot for $\Omega_{\rm{DE}}$ vs $z$. using positive (\emph{left}) and negative (\emph{right}) values for the parameter $b$}
    \label{fig4}
\end{figure*}

In Fig. \ref{fig2} we regard the evolution of the deceleration parameter $q$ as function of the redshift $z$ and for positive and negative values of the parameter $b$. In this case, we can see that in both cases, as $b$ decreases, $q$ progressively converges towards the $\Lambda$CDM model. In Fig.~\ref{fig3} (left panel) we depict the cosmological evolution of $w_{\rm{eff}}$ as a function of the redshift $z$ and $b$ (positive and negative values), and using some fixed positive values of $b$ (right panel). 	We can see that, independently on the choice of $b$, $w_{\rm{eff}}$ starts from zero in the matter dominated era and asymptotically approaches -1 (regardless of the sign of $b$, $w_{\rm{eff}}$ shows quintessence like behavior). The deviation from $\Lambda$CDM model prediction is larger as $b$ increases. A similar behavior can be appreciated in both figures of Fig.~\ref{fig4}. 

In general, from above we can deduce that as the magnitude of $b$ increases, the present model deviates from $\Lambda$CDM model. This behavior is the expected, since by construction our approximated solution for $H(z)$ is built as a perturbation of the $\Lambda$CDM model solution

\section{Parameter constraints}\label{params} 
To constraint the parameters of the model we implement a statistical analysis based in the Markov Chain Monte Carlo (MCMC) method, considering the observational data of the Hubble parameter, $H_{{\rm obs}}(z)$, obtained from the cosmic chronometers and the radial BAO size methods (see Ref.~\cite{Cao:2021uda} for the complete list of data points and information about the  observational methods), and of the magnitude-redshift relation, $m_{{\rm obs}}(z)$, for the Pantheon sample including 1048 observational data \cite{Efstathiou:2021ocp}.

As exposed by Eq.~(\ref{eq_fQmodel}), our model depends on two dimensionless parameters ($n, b$); however, we are considering the specific case of $n = 1$. Then, for the statistical analysis presented in this section, we also decided to include the current value of the Hubble parameter, $H_0 \equiv H(z = 0)$, as well as the current value of the matter density parameter, $\Omega_{m0} \equiv \Omega_m(z = 0)$. With this, we compute the value of the Hubble parameter at different redshifts as predicted by the model, $H_{{\rm fQ}}(z)$, as well as the apparent magnitude $m_{{\rm fQ}}(z)$ (which is a function of the expansion rate, $H_{{\rm fQ}}(z)$ through the luminosity distance; see \cite{Efstathiou:2021ocp} for details), considering the approximate solution given by Eq.~(\ref{eq25}).

We implement the MCMC method by using the 
\texttt{emcee} package \cite{Foreman-Mackey:2013}, which employs a suitable $\chi^2$ function as an input for the applied Maximum Likelihood (ML) method. In our case, the likelihood function reads
\begin{equation}\label{eq_likelihood}
    \mathcal{L}(D|z;\bm{\theta}) = -\ln p(D|z;\bm{\theta}) = -\frac{1}{2} \chi^2(z;\bm{\theta}),
\end{equation}
with the $\chi^2$ function defined as
\begin{equation}\label{eq_chisq}
    \chi^2(z;\bm{\theta}) = \sum_{k=1}^{N_d}
    \frac{\left(
    D_{{\rm fQ},k}(z;\bm{\theta}) - D_{{\rm obs},k}(z)
    \right)^2}{\sigma_k^2},
\end{equation}
where $\bm{\theta} = \left(H_0,\Omega_{m0},b\right)$ is the vector of the parameters to be fit, $\sigma_k$ is the standard deviation of the corresponding data, and $N_d$ is the number of data points. Depending on the set of observational data, $D_{{\rm fQ}}$ corresponds to the expansion rate, $H(z)$ (with $N_d = 40$), or to the apparent magnitude, $m$ (with $N_d = 1048$).

The final important aspect to mention here is the introduction of appropriate priors for the parameters, which are displayed in Tab.~\ref{tab_priors} and are used for both of the analyses. Notice that for $H_0$ the prior is uniform and large enough for the MCMC to be able to explore values around the two reported observations which are currently under scrutiny (see a discussion bellow).
\begin{table}[ph]
\centering
\caption{Intervals used as the (flat) priors for the parameter studied in this analysis, $\left(H_0,\Omega_{m0},b\right)$.}
{\begin{tabular}{@{}cccc@{}} \hline
       $\theta_i$ & $H_0$ (km/s/Mpc) & $\Omega_{m0}$ & $b$ \\ \hline
       Prior      & (60, 80) & (0.1, 0.4)    & (-2, 2) \\ \hline
\end{tabular}
\label{tab_priors}}
\end{table}

\begin{figure*}
\centering
    \includegraphics[width=1.0\textwidth]{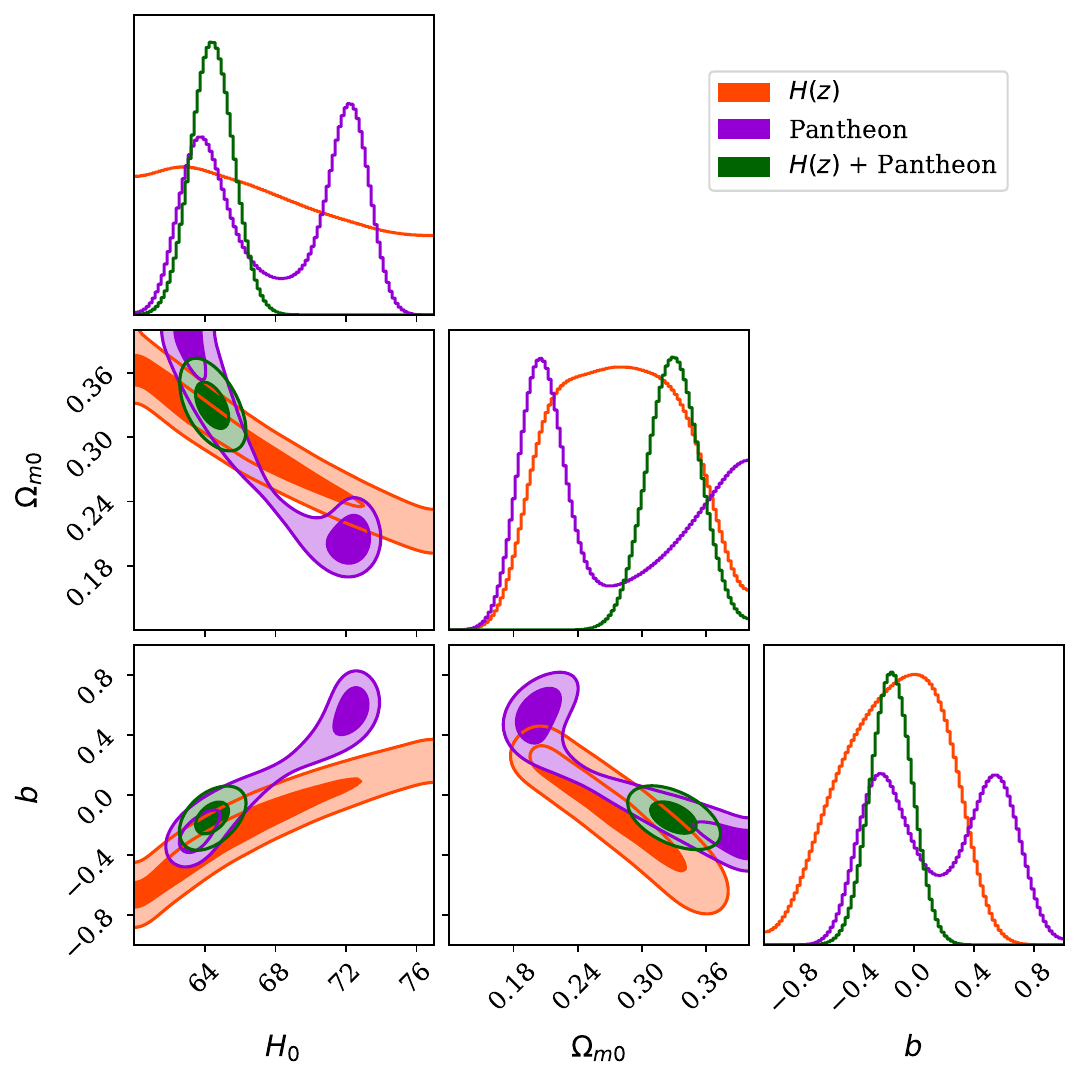}
    \caption{The result of the MCMC analysis for the $H(z)$ observational data (orange), the Pantheon sample (purple), and the combination of both datasets (green). The $1-$ (histograms) and $2-$D (contours) projections of the posterior probability distributions are shown for the model parameters $\left(H_0,\Omega_{m0},b\right)$.}
    \label{fig_cornerCombo}
\end{figure*}

\begin{figure*}
\centering
    \includegraphics[width=1.0\textwidth]{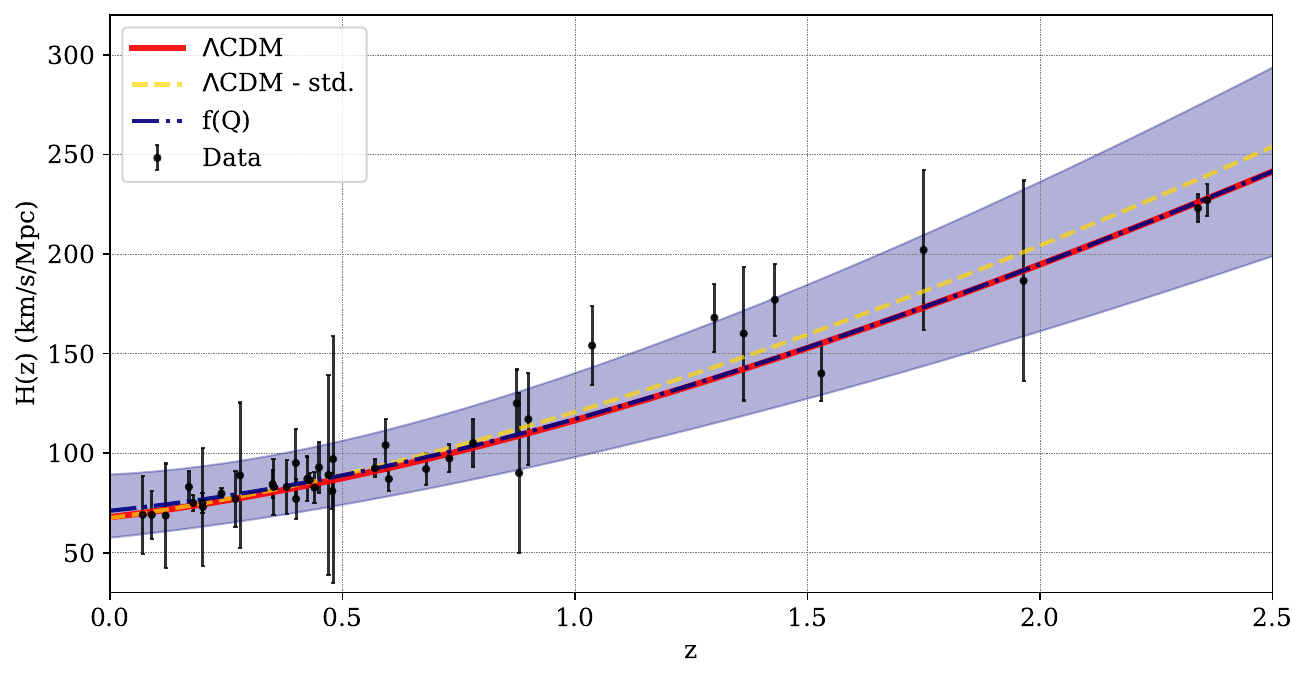}
    \includegraphics[width=1.0\textwidth]{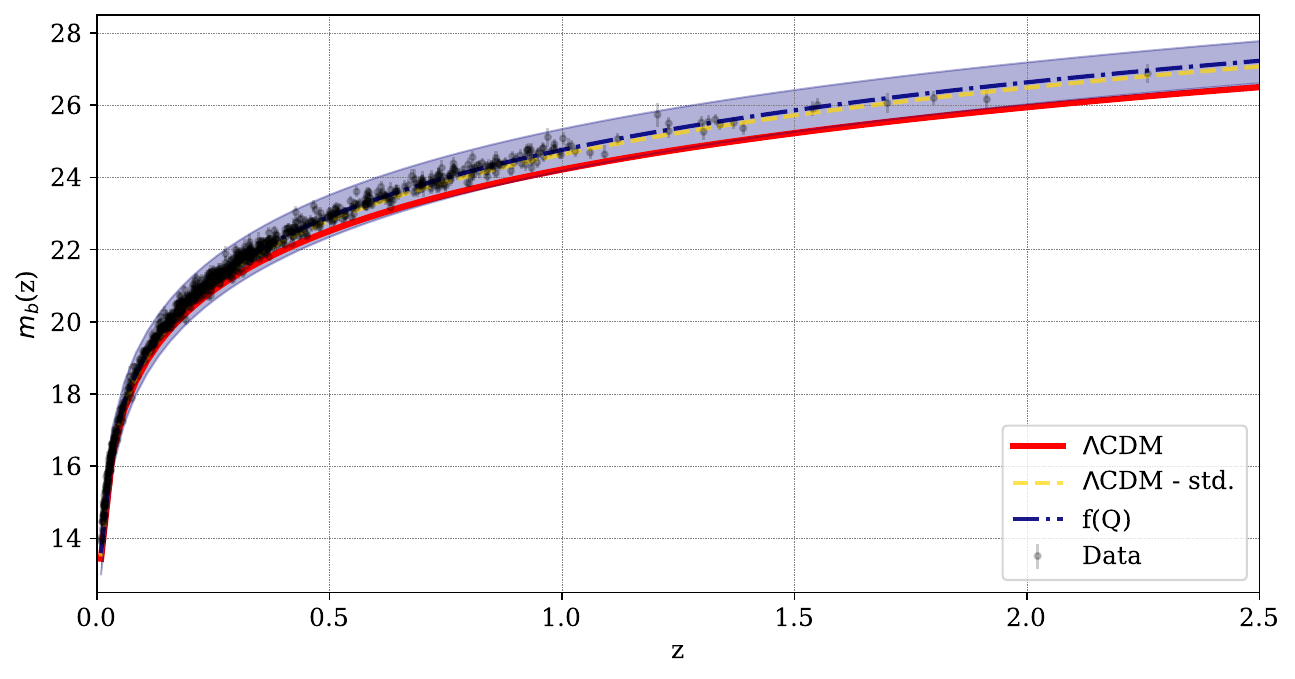}
    \caption{Evolution of the Hubble parameter (top) and the SN magnitude (bottom) with the redshift, $z$, as predicted for the $f(Q)$ model presented in this work (dot-dashed blue), compared against observational data (black dots with the vertical lines indicating the uncertainty). The predicted evolution by the $\Lambda$CDM (full red and dashed gold) model is also shown for comparison, as detailed in the text.}
    \label{fig_compare_Hz}
\end{figure*}

After running \texttt{emcee} with 100 walkers and 40000 steps (for each data set), the results of our MCMC analyses are presented in Fig.~\ref{fig_cornerCombo}, where the $1-$ and $2-$D projections of the posterior probability distributions of the studied parameters are shown. In particular, the $2-$D plots include the $1\sigma$ and $2\sigma$ C.L.~contours for each parameter. An important and remarkable characteristic of these plots is that one can notice the strong correlation among the parameters: both couples, $\left(H_0,\Omega_{m0}\right)$ and $\left(\Omega_{m0},b\right)$ have a manifest negative correlation, while for $\left(H_0,b\right)$, the correlation is clearly positive (all this inside the corresponding studied regions). These correlations emphasize the significant impact of the perturbation parameter $b$, which when different from zero, allows the model to fit the data with appropriate values of the other parameters (see Table \ref{tab_MCMCintervals} and Fig.~\ref{fig_compare_Hz}). 
\begin{table}[ph]
\centering
\caption{Resulting $1\sigma$ allowed intervals for the fitted parameters from the MCMC analysis. The values for the $\Lambda$CDM model are from Planck 2018 \cite{Planck:2018vyg}.}
{\begin{tabular}{@{}cccc@{}} \hline
        Model & $H_0$  (km/s/Mpc) & $\Omega_{m0}$ & $b$ \\ \hline
        $\Lambda$CDM & $67.36\pm0.54$ & $0.315\pm0.007$ & -- \\
        $f(Q): H(z)$ & $67.79_{-5.35}^{+7.56}$ & $0.279_{-0.060}^{+0.059}$ & $-0.116_{-0.362}^{+0.309}$ \\ 
        $f(Q):$ Pantheon & $68.25_{-4.80}^{+4.22}$ & $0.238_{-0.038}^{+0.135}$ & $0.163_{-0.430}^{+0.425}$ \\ 
        $f(Q): H(z)+$Pantheon & $64.41_{-0.61}^{+0.67}$ & $0.330\pm0.018$ & $-0.151_{-0.069}^{+0.068}$ \\ \hline
        \end{tabular}
    \label{tab_MCMCintervals}}
\end{table}

Interestingly, the results of the SN data (purple lines and contours in Fig.~\ref{fig_cornerCombo}) appear to indicate that our model fits the data well with either of the two values of $H_0$ that have been published by two independent collaborations ($H_0^{{\rm Plank}} = (67.36\pm 0.54)\, \rm{km}\,\rm{s}^{-1}\,\rm{Mpc}^{-1}$ by Planck 2018 \cite{Planck:2018vyg}, and $H_0^{{\rm SH0ES}} = (73.30 \pm 1.04)\, \rm{km}\,\rm{s}^{-1}\,\rm{Mpc}^{-1}$ by SH0ES using the Cepheid-calibrated cosmic distance ladder \cite{Riess:2021jrx}\footnote{Notice that the latest measurement by SH0ES is $H_0 = (73.01 \pm 0.99)\, \rm{km}\,\rm{s}^{-1}\,\rm{Mpc}^{-1}$ and $H_0 = (73.15 \pm 0.97)\, \rm{km}\,\rm{s}^{-1}\,\rm{Mpc}^{-1}$, respectively, a $5\%$ or $7\%$  reduction in the uncertainty  \cite{Riess:2022mme}.}), while the expansion rate data exhibits a weak preference for $H_0^{{\rm Plank}}$. However, it is clear that the combination of the two datasets indicates a strong preference for the lower value, with the appropriate combination of the other parameters involved.

Now, as one would expect, the constraints on the parameters is stronger when the two datasets are combined in a joint analysis using 
\begin{equation}\label{eq_chi2Combo}
    \chi^2(z;\bm{\theta}) = \sum_{k=1}^{40}
    \frac{\left(
    H_{{\rm fQ},k}(z;\bm{\theta}) - H_{{\rm obs},k}(z)
    \right)^2}{\sigma(H)_k^2} 
    +
    \sum_{k=1}^{1048}
    \frac{\left(
    m_{{\rm fQ},k}(z;\bm{\theta}) - m_{{\rm obs},k}(z)
    \right)^2}{\sigma(m)_k^2},
\end{equation}
in Eq.~(\ref{eq_likelihood}). The results of running \texttt{emcee} in this case are presented in the fourth row of Table \ref{tab_MCMCintervals}, and in the corresponding (green) contours and histograms in Fig.~\ref{fig_cornerCombo}. As discussed before regarding the prediction for $H_0$, in this case the model prefers a value which is closer to the one measured by Planck, and to make a better fit to the data, $\Omega_{m0}$ (within its $1\sigma$ C.L.) is found to be compatible with the $\Lambda$CDM prediction. Also, the perturbation parameter $b$ is found to be negative, excluding $b=0$ at $> 2\sigma$ C.L.

The top panel of Fig.~\ref{fig_compare_Hz} shows the evolution of the Hubble parameter as a function of the redshift, comparing the prediction by the $\Lambda$CDM model, the $f(Q)$ (dot-dasshed--blue line) presented in this work, and the observational data. For the $\Lambda$CDM prediction, we included two lines for a better comparison with our model: the full--red line is the prediction obtained setting $H_0$ and $\Omega_{m0}$ to the values obtained from the present MCMC analysis explained above (see the second row of Table \ref{tab_MCMCintervals}); on the other hand, the dashed--gold line (labeled as $\Lambda$CDM - std.) is obtained from the standard prediction of the $\Lambda$CDM model, i.e., using the parameters of the first row of Table \ref{tab_MCMCintervals}.

Similarly, the bottom panel of Fig.~\ref{fig_compare_Hz} presents the same comparison for the apparent magnitude. In this case, the $f(Q)$ (dot-dasshed--blue) and $\Lambda$CDM (full--red) predictions are drawn considering the corresponding parameters in the third row of Table \ref{tab_MCMCintervals} (with $b=0$ for the later). The deviation of the $\Lambda$CDM (full--red) from the data is apparent and expected considering that both, $H_0$ and $\Omega_{m0}$, are considerably different from those predicted by the stantdard $\Lambda$CDM model (dashed--gold).

The similarity of our $f(Q)$ model with the prediction of $\Lambda$CDM is evident. When the same values for the shared parameters ($H_0$ and $\Omega_{m0}$) are used, the models deviate from each other, specifically regarding the apparent magnitude prediction; as a matter of fact, our model fits better those data than $\Lambda$CDM (red line in Fig.~\ref{fig_compare_Hz}). Comparing the proposed $f(Q)$ against the standard $\Lambda$CDM (gold-dashed line in Fig.~\ref{fig_compare_Hz}), we see that, as explained in Sec.~\ref{cosmo-analysis}, the former can be interpreted as a perturbation of the later.

\begin{figure*}
\centering
    \includegraphics[width=1.0\textwidth]{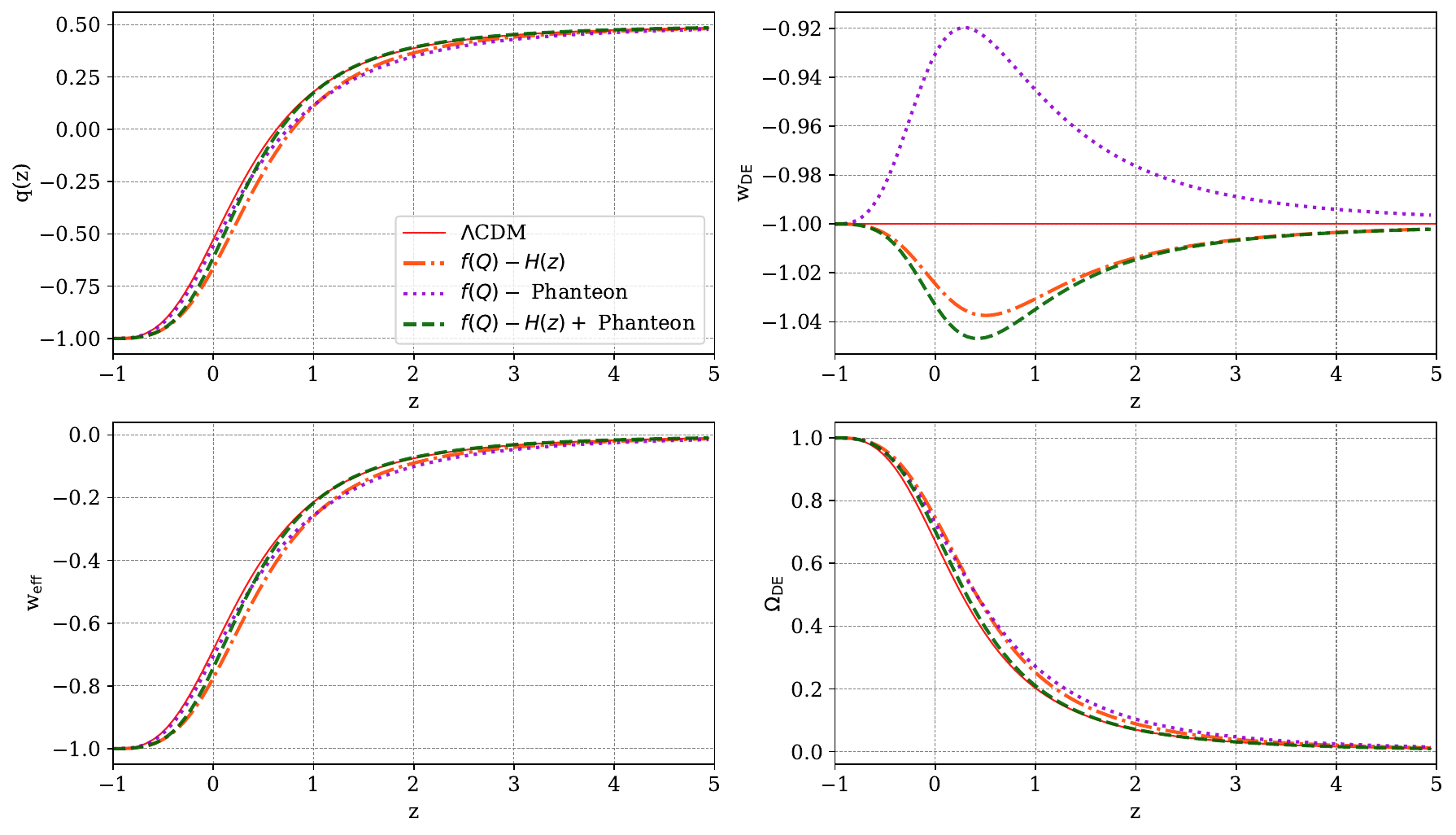}
    \caption{Evolution of some cosmological parameters vs.~the redshift, $z$, as predicted for the $f(Q)$ model presented in this work, compared with the $\Lambda$CDM predictions (red-full line). In each panel, the dot-dashed orange, dotted violet and dashed green lines are the result of the parameter constraints from $H(z)$, Pantheon and $H(z) + $ Pantheon samples, respectively.}
    \label{fig_compare_cosmoParams}
\end{figure*}
\begin{table}
\centering
\caption{Present ($z=0$) value of $w_{\rm{DE}}$ and $\Omega_{\rm{DE}}$ as obtained from the proposed model, compared with the Planck 2018 results \cite{Planck:2018vyg}.}
{\begin{tabular}{@{}ccc@{}} \hline
    Model                 & $w_{\rm{DE}0}$  & $\Omega_{\rm{DE}0}$ \\ \hline
    $\Lambda$CDM          & $-1.03\pm 0.03$ & $0.6847\pm0.0073$   \\
    $f(Q): H(z)$          & $-1.02$         & $0.7458$            \\ 
    $f(Q):$ Pantheon      & $-0.93$         & $0.7280$            \\ 
    $f(Q): H(z)+$Pantheon & $-1.03$         & $0.7029$            \\ \hline
\end{tabular}
\label{tab_PlanckCompare}}
\end{table}
This is also discernible by looking at the evolution of the cosmological parameters $w_{{\rm DE}}$, $w_{{\rm eff}}$, $\Omega_{{\rm DE}}$ and the deceleration parameter, $q$, as shown in Fig.~\ref{fig_compare_cosmoParams}. Again, the predictions of the $f(Q)$ model for each of the data analysis presented above are compared with $\Lambda$CDM. Overall, for $q$, $w_{{\rm eff}}$, and $\Omega_{{\rm DE}}$, the combined data analysis result (green-dashed line) is the closest to $\Lambda$CDM (red-full line). The plot of of $w_{{\rm DE}}$ (top right panel) demonstrates the important difference of the model when the perturbative parameter $b$ is positive or negative. While the outcome of the fit to the Pantheon sample (dotted-violet line) is such that $b>0$ (quintessence-like), the other two analyses ($H(z)$ data alone and $H(z)$+Pantheon) give $b<0$ (phantom-like), though the deviation from $\Lambda$CDM is less than 10\%. A comparison of the present value (at $z=0$) of the DE equation of state and the DE density predicted by the proposed model (using the best fit parameters shown in Table \ref{tab_MCMCintervals}) with the Planck 2018 results \cite{Planck:2018vyg} (see Table \ref{tab_PlanckCompare}), further support this: constraints from the combination of the two data sets predict values closer to $\Lambda$CDM, while it appears that the fit to the Pantheon data set would predict a larger value of $w_{\rm{DE}0}$ ($>-1$), compared against $\Lambda$CDM.

\section{Conclusions}\label{conclus} 
Until now, a definitive and satisfactory resolution to the enigma of DE remains elusive within the cosmological community. Consequently, cosmologists continue to propose new and often exotic theories in their pursuit of someday attaining a comprehensive explanation for this perplexing problem. Among these proposals, the $f(Q)$ non-metric gravity has recently gained prominence as an alternative framework for elucidating phenomena in both early and late-time cosmology. Remarkably, this theory achieves this without necessitating the inclusion of dark energy, the inflaton field, or dark matter. This innovative approach represents a significant step forward in our quest to understand the fundamental nature of the universe. 

In this sense, here we have introduce 
a new and viable $f(Q)$ gravity model which can be represented as a perturbation of the $\Lambda$CDM model (see Eq.~(\ref{eq_fQmodel})). Usually, within the realm of $f(Q)$ gravity, the customary approach to investigate cosmological evolution involves employing a parametrization of the Hubble parameter in terms of redshift, among other strategies. In this work 
as an alternative, we have derived an analytical approximation for the expansion rate $H(z)$ (Eq.~(\ref{eq25})), and from it, we have deduced approximated analytical expressions for the parameters $w_{\rm{DE}}$, $w_{\rm{eff}}$, and $\Omega_{DE}$, as well as the deceleration parameter $q$ in terms of the redshift $z$ (Eqs.~(\ref{eq30})-(\ref{eq33})). 

In order to verify the viability of this approximate analytical solution for $H(z)$,  we have examined the behaviour of the these parameters in the late-time regime. It should be noted that for $b>0$, $w_{\rm{DE}}$ shows a quintessence like behavior and for $b<0$, it shows a phantom like behavior (see Fig.~\ref{fig1}). However, regardless of the sign of $b$, $w_{\rm{eff}}$ shows a quintessence like behavior (shown in Fig.~\ref{fig3}). Furthermore, it has been deduced that in general as the magnitude of the parameter $b$ increases, the present model deviates progressively from the $\Lambda$CDM model (depicted in Figs.~\ref{fig1}-\ref{fig4}). This is the expected behavior, since by construction our approximated solution for $H(z)$ was built as a perturbation of the $\Lambda$CDM model solution. 

The MCMC analysis considering the Hubble parameter and the Pantheon data samples allowed us to find constraints on the perturbation parameter $b$, as well as on the current values of the Hubble expansion rate ($H_0$) and the matter density ($\Omega_{m0}$), as presented in Table \ref{tab_MCMCintervals} and Fig.~\ref{fig_cornerCombo}. The combined datasets ($H(z) +$ Pantheon) analysis indicates a preference for negative values of $b$, implying a phantom-like behavior for $w_{{\rm DE}}$, and making evident that the model deviates only marginally from the $\Lambda$CDM predictions (see Figs.~\ref{fig_compare_Hz} and \ref{fig_compare_cosmoParams}), a conclusion which is also supported by the predicted values of the DE equation of state ($w_{\rm{DE}0}$) and density ($\Omega_{\rm{DE}0}$), at present times ($z=0$), as reported in Table \ref{tab_PlanckCompare}. Then, our findings indicate that this $f(Q)$ gravity model is indeed a viable candidate for describing the late-time evolution of the Universe at the background level.


\noindent \textbf{Acknowledgements}
A. O. is supported by Patrimonio Aut\'onomo-Fondo Nacional de Financiamiento para la Ciencia, la Tecnolog\'ia y la Innovaci\'on Francisco Jos\'e de Caldas (MINCIENCIAS-COLOMBIA) Grant No. 110685269447 RC-80740-465-2020, projects 69723 and 69553. We also thank Ricardo Vega (Universidad del Atlántico) for helping us with the computer resources of his Laboratory used for the MCMC analyses.

\end{document}